\documentclass[aps,prb,preprint,amsfonts,amsmath,amssymb,floatfix,10pt,reprint]{revtex4-1}
\usepackage{braket} 
\usepackage{graphicx}

\usepackage{caption}

\usepackage{algorithm} 
\usepackage{algpseudocode}
\floatname{algorithm}{Algorithm}

\usepackage{hyperref}

\usepackage{color}
\usepackage[normalem]{ulem}

\usepackage[bottom]{footmisc}
\algnewcommand{\LineComment}[1]{\State\(\triangleright\) #1}

\graphicspath{{figures/}}

\begin{document}
\title{Simple heuristics for efficient parallel tensor contraction and quantum circuit simulation}

\author{Roman Schutski}
\affiliation{Center for Computational and Data-Intensive Science and Engineering, Skoltech, Skolkovo Innovation Center, Moscow Region, 121205, Russian Federation}

\author{Dmitry Kolmakov}
\affiliation{Central Research Institute, Huawei Technologies}

\author{Taras Khakhulin}
\affiliation{Center for Computational and Data-Intensive Science and Engineering, Skoltech, Skolkovo Innovation Center, Moscow Region, 121205, Russian Federation}

\author{Ivan Oseledets}
\affiliation{Center for Computational and Data-Intensive Science and Engineering, Skoltech, Skolkovo Innovation Center, Moscow Region, 121205, Russian Federation}
\date{\today}

\begin{abstract}
Tensor networks are the main building blocks in a wide variety of computational sciences, ranging from many-body theory and quantum computing to probability and machine learning. Here we propose a parallel algorithm for the contraction of tensor networks using probabilistic graphical models. Our approach is based on the heuristic solution of the $\mu$-treewidth deletion problem in graph theory. We apply the resulting algorithm to the simulation of random quantum circuits
and discuss the extensions for general tensor network contractions.
\end{abstract}


\maketitle

\section{{Introduction}
\label{sec:introduction}}
In recent years we have witnessed an explosive growth in the numerical techniques for solving high-dimensional problems. Substantial understanding and quantitative accuracy of the simulation were reached in many-body physics.\cite{evenbly2011tensor} At the same time, the superhuman performance was achieved by neural networks in solving extremely high dimensional problems, like image and speech recognition or complex games.\cite{radford2018improving, silver2017mastering} The ever-increasing need for computational resources stimulates the research into novel computing devices, such as quantum computers.~\cite{IntelQ, IBMQ, arute2019quantum} We would speculate that a significant part of the mentioned advances is due to the establishment of tensor networks as a universal language for high-dimensional modeling, and the development of efficient tools to manipulate them.

Tensor networks were proposed for the simulation of quantum circuits by Markov and Shi.\cite{markov2008simulating} The authors showed that the evaluation of a quantum circuit on a classical computer amounts to the contraction of the corresponding tensor network. The graph-based notation employed by Markov and Shi has been well established in many-body physics by the time of their work.~\cite{bridgeman2017hand} Following the original work, several authors proposed highly efficient algorithms for quantum circuit simulation, see~\onlinecite{chen201864, pednault2017breaking, li2018quantum, gray2020hyper} for more details. Tensor contraction algorithms were also studied in many-body physics context.\cite{pfeifer2014faster}

Boixo et al. did the next important step in understanding tensor contractions in quantum circuit simulation context.~\cite{Boixo2017} The authors proposed to use graphical models to represent tensor networks, which are line graphs of the traditional circuit notation. Following Boixo et al., the contraction of a network amounts to the Belief propagation or Bucket elimination\cite{detcher2013bucket} procedure developed in statistics. In addition to establishing a link with statistical analysis, graphical models avoid the use of hypergraphs, which are necessary in the traditional representation of tensor networks.~\cite{pednault2017breaking} In our recent work\cite{schutski2019adaptive}, we proposed an algorithm for \emph{partial} tensor network contraction in the graphical model notation.

In this article, we explore algorithms for \emph{parallel} tensor contraction and quantum circuit simulation. First, we describe a general parallel algorithm, similar to the one proposed by Chen et al.~\cite{Chen2018}. As in any parallel algorithm, the algorithm in ~\onlinecite{Chen2018} splits the initial circuit simulation task into multiple subtasks, which can be evaluated independently. The core step of this algorithm depends on the choice of the subtasks. Here we show how to select subtasks to achieve maximal computational efficiency. As with finding an optimal way to contract an arbitrary tensor network,~\cite{markov2008simulating} finding an optimal parallelization scheme implies solving an NP-complete problem.\cite{fomin2012planar} We propose a simple yet very efficient heuristic to find the parallelization scheme. We also present the extension of our techniques to general tensor networks, e.g., not necessarily the ones associated with quantum circuits.

The paper is organized as follows. First, we briefly review the use of graphical models to represent tensor contractions in Section \ref{sec:intro_models}. It is well known \cite{markov2008simulating, Boixo2017, schutski2019adaptive} that the order of contraction of the network dramatically influences the numerical cost of this operation. Finding an optimal order (e.g., the one with the lowest cost) amounts to finding an optimal \emph{tree decomposition} (TD) of the graphical model, an NP-complete problem. We review the connection between the orderings and tree decompositions in Section \ref{ssec:tree_decomposition}. The characteristic of the tree decomposition called \emph{treewidth} defines the numerical complexity of the contraction of the tensor network. An efficient parallelization procedure thus has to split the full contraction task into subtasks with minimal treewidth. We propose several ideas to implement simple yet very efficient heuristics to achieve this in Section \ref{sec:heuristics}. We verify our findings with numerical experiments in Section \ref{sec:experiments}. The outlook is provided in Section \ref{sec:outlook}.

\section{Graphical models for tensor contraction and quantum circuit simulation \label{sec:intro_models}}
\subsection{Graphical models\label{ssec:graphical_models}}
In this section we review the use of graphical models for the representation of tensor networks. For a more extensive introduction in the scope of quantum circuit simulation, the readers are referred to previous works \cite{markov2008simulating, Boixo2017, schutski2019adaptive}. Here we give a formulation for general tensor networks.

\begin{figure}
\centering
\includegraphics[width=0.25\textwidth]{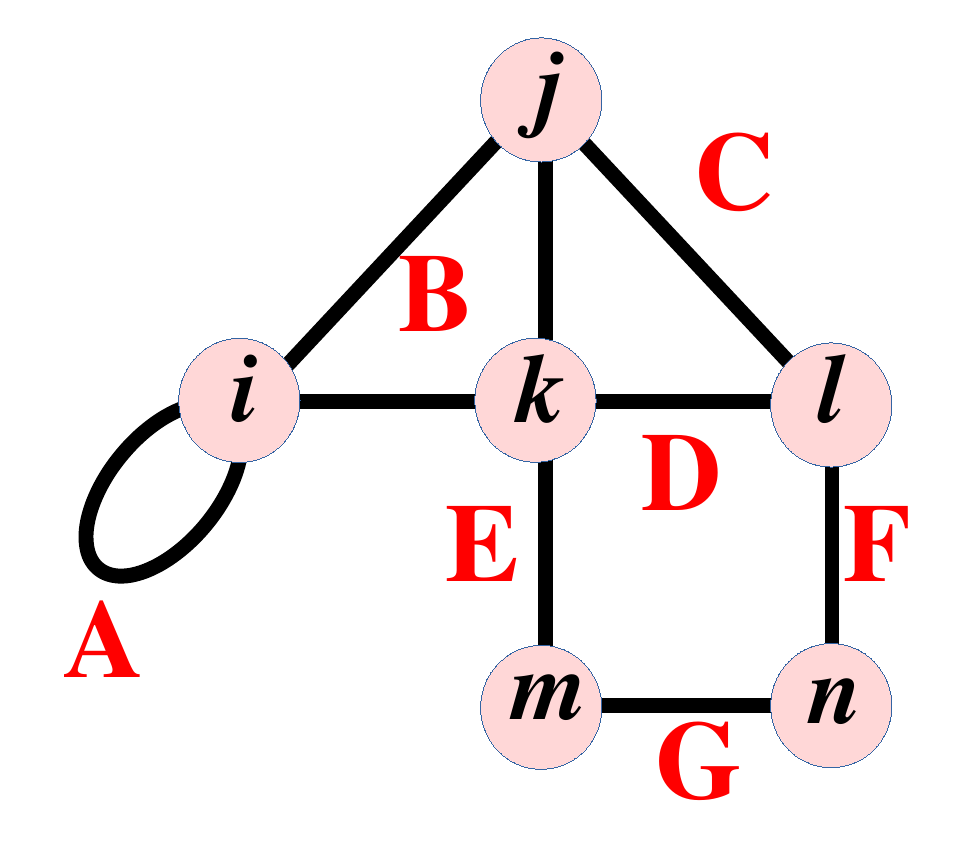}
\caption{Representation of a tensor network in Eq.~\ref{eq:tensor_network} by a graphical model.\label{fig:tensor_network}}
\end{figure}

A tensor network is a product of tensors (multidimensional arrays). We will use capital letters $A, B, C\dots$ to represent tensors and lowercase letters $i, j, k\ldots$ to denote indices and scalars. The main operation on tensor networks is contraction, e.g. a summation over a subset of indices. A toy example of a tensor network is given in Eq.\ref{eq:tensor_network} (product of all terms is assumed):
\begin{equation}
    A_{i} B_{ijk} C_{jl} D_{kl} E_{km} F_{ln} G_{mn}
\label{eq:tensor_network}
\end{equation}
This network can be represented by the graph in Fig.~\ref{fig:tensor_network}.

Quantum circuits can be readily represented by the graphs analogous to Fig.~\ref{fig:tensor_network}.\cite{Boixo2017, schutski2019adaptive} Note that this notation is essentially the same as the one used for Bayesian networks and Markov random fields. The nodes here represent indices of the expression, and tensors are reflected by cliques (fully connected subgraphs) in the expression's graph (also shown in red in Fig.~\ref{fig:tensor_network}). We denote single index tensors by self-loops and omit parallel edges (formally we have to use multigraphs in the notation, but this detail does not affect further discussion).

\begin{figure}
\centering
\includegraphics[width=0.49\textwidth]{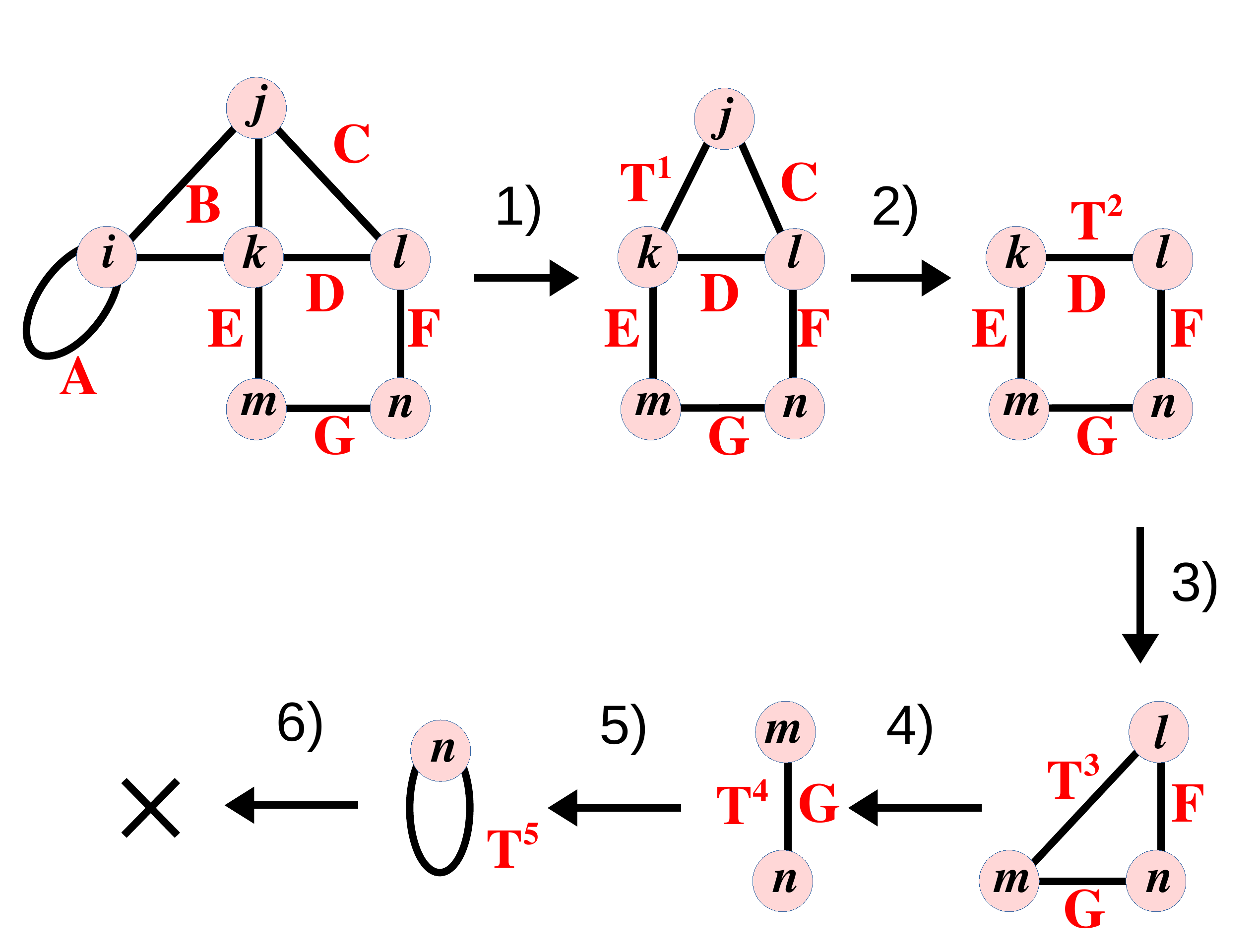}
\caption{Contraction of a tensor network in Eq.~\ref{eq:tensor_network}.\label{fig:sequence}}
\end{figure}

Let us now contract the network in Eq.~\ref{eq:tensor_network}. Assuming that the dimension of every index is $L$ (for quantum circuit simulation $L=2$), this contraction can be evaluated using the following sequence of operations (we specify the scaling of the number of operations to the right of each step):
\begin{equation}
    \begin{split}
        & \sum_{ijklmn} A_{i} B_{ijk} C_{jl} D_{kl} E_{km} F_{ln} G_{mn} = \sigma \\
        & 1) ~\sum_{i} A_{i} B_{ijk} = T^{1}_{jk} \quad \mathcal{O}(L^3) \\ 
        & 2) ~\sum_{j} C_{jl} T^{1}_{jk} = T^{2}_{kl} \quad \mathcal{O}(L^3) \\
        & 3) ~\sum_{k} D_{kl} T^{2}_{kl} E_{km} = T^{3}_{ml} \quad \mathcal{O}(L^3) \\
        & 4) ~\sum_{l} F_{ln} T^{3}_{ml} = T^{4}_{nm} \quad \mathcal{O}(L^3) \\
        & 5) ~\sum_{m} T^{4}_{nm} G_{nm} = T^{5}_{n} \quad \mathcal{O}(L^{2}) \\
        & 6) ~\sum_{n} T^{5}_{n} = \sigma \quad \mathcal{O}(L)
    \end{split}
    \label{eq:sequence}
\end{equation}
The contraction sequence can also be conveniently represented by graphical models, as shown in Fig.~\ref{fig:sequence}.
At each step of contraction a vertex is removed from the expression graph and all of its neighbors are connected into a new clique, which corresponds to an intermediate tensor (denoted by $T$'s in Fig.~\ref{fig:sequence}). The size of the clique is the dimension of the intermediate. The exponent in the numerical cost of the contraction is greater by one than the size of the intermediate (assuming all indices have the same size).

Graphical models can be used to find optimal contraction sequences of tensor networks. Assume we would like to contract a given tensor network with a minimal number of multiplications and additions. We would then need to find a sequence of node eliminations, such that the maximal size of the cliques in the sequence (and hence the dimension of the intermediates) is minimized. This problem is NP-hard~\cite{bodlaender1994tourist, blair1993introduction} and amounts to finding a TD of the expression's graph (however, many efficient algorithms exist which can calculate close to optimal solutions, see recent examples in Refs. \onlinecite{gogate2004complete, tamaki2019positive, strasser2017computing}). We will explain the TD and its relation with elimination orders in the following sections. The size of the maximal clique in the contraction sequence is the treewidth; 
we are interested in finding orders corresponding to minimal treewidth.

\subsection{Tree decompositions\label{ssec:tree_decomposition}}
In this section, we will formally define a standard TD and relate this concept with the elimination orders. This relation of linear orderings of graph vertices with tree graphs will be employed to construct efficient parallelization heuristics.

Tree decompositions were introduced by Robertson and Seymour\cite{robertson1986graph}; the reader is referred to \onlinecite{bodlaender1994tourist, blair1993introduction} for an alternative introduction to the topic. All graphs in this section are simple and undirected if not stated otherwise (which means they do not contain self-loops and parallel edges \footnote{Simple graphs do not restrict the analysis of computational complexity of tensor network contractions, as any tensor network can be transformed such that its expression graph is simple. To achieve this, one needs to multiply factors on parallel edges and to contract self-loops. These operations do not significantly increase numerical complexity}).

We start with a graph $G = (V, E)$ where $V$ is the set of vertices, and $E$ is the set of edges. Tree decomposition is a mapping of the initial graph $G$ into a tree graph $F = (B, T)$, where $B$ is the set of \emph{bags} (nodes) and $T$ are the edges of the tree. Each bag $b \in B$ is a subset of nodes of the initial graph $G$, e.g., $b \in V$. A TD has to fulfill three criteria to be correct:
\begin{enumerate}
\item Every node is in some bag, i.e., $\cup_{b \in B} b = V$.
\item For every edge $(u, v) \in E$ there must be a bag such that both endpoints are in that bag, i.e., $\exists b: u \in b, v \in b$.
\item For every node $u$ of $G$, the subgraph of the tree $F$, induced by all bags that contain $u$ is a connected tree.
\end{enumerate}

\begin{figure}[ht!]
\centering
\includegraphics[width=0.40\textwidth]{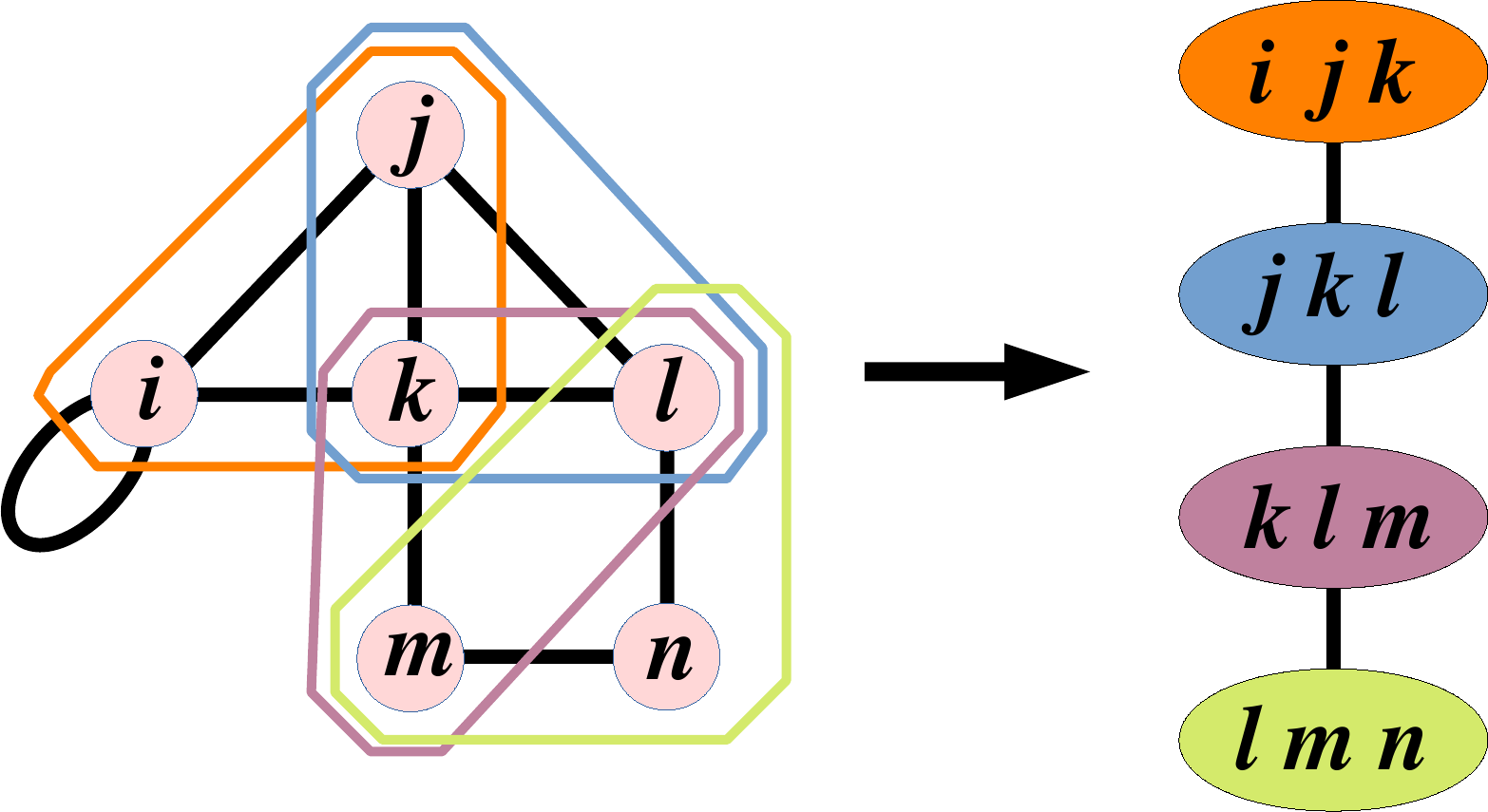}
\caption{Graphical model and its tree decomposition (in this particular case the tree is a path graph). The decomposition corresponds to the order $[i ~ j ~ k ~ l ~ m ~ n]$. 
\label{fig:tree_decomposition}}
\end{figure}

The width of TD is the maximal size of the bag minus one. Informally, the treewidth quantifies how much a given graph resembles a tree; the treewidth of trees is one. Bags in tree decomposition are exactly cliques that are formed in some contraction sequence (these cliques include the node which is eliminated at each step). Finding a TD of minimal width in NP-complete.\cite{} An example of TD is shown in Fig.~\ref{fig:tree_decomposition} and is further explained below.


The TD of the graph in Fig.~\ref{fig:tensor_network} which corresponds to the sequence $\pi = [i ~ j ~ k ~ l ~ m ~ n]$ is shown in Fig.~\ref{fig:tree_decomposition}. We have to note, however, that the map between the orderings of vertices and tree graphs in not bijective: multiple orderings can correspond to the same TD. For example, the order $\tilde{\pi} = [n ~ m ~ l ~ k ~ j ~ i]$ yields the same tree. An algorithm for building a tree decomposition for a given elimination order is provided in Appendix~\ref{sec:building_tree_decomposition}. The reverse operation, e.g. a procedure to obtain \emph{some} elimination order for a given tree $F$, is provided in the Appendix~\ref{sec:finding_orderings}.


\subsection{Determining treewidth\label{sssec:determining_treewidth}}
For completeness, we briefly explain the procedure to calculate the treewidth provided an elimination order or a TD. This simple procedure provides a way to estimate the quality of different TDs/contraction sequences of a given tensor network. We use it to compare performance of different parallelization algorithms.

Given an elimination order $\pi$, a corresponding treewidth $\tau$ is calculated by performing the elimination procedure and finding the size of the maximal clique which will emerge during this process. Notice that this operation is linear in the size of the graph (in contrast with finding the order with the smallest treewidth, which is NP-complete). The algorithm is summarized in Alg.~\ref{code:find_treewidth_order}.

\begin{algorithm}[H]
  \caption{Finding treewidth from the elimination order}
  \label{code:find_treewidth_order}
  \begin{algorithmic}[1]
    \Require $G = (U, E)$, $\pi = \{(u_{i}, i)\}_{i=1}^{|U|}$
    \Ensure $\tau$
    \Statex
    \Function{Find\_treewidth\_from\_order}{$G, \pi$}
    \State $\tau \gets 1$
    \For{$u \in \pi$}\Comment{Eliminate according to the order}
        \For{$(v, w) \in \mathcal{N}(u)$}
            \State $E \gets E \cup (v, w)$
        \EndFor
        \State $\tau \gets \max(|\mathcal{N}(u)|, \tau)$
        \State $U \gets U \setminus u$
    \EndFor
    \EndFunction
  \end{algorithmic}
\end{algorithm}

Alternatively, if the tree decomposition $F$ of $G$ is provided, then the treewidth is the size of the maximal bag in $F$ minus $1$: $\tau = \underset{b \in F}{\max} |b| - 1$. It is again apparent that the cost of the determination of treewidth of a tree $F$ is linear in the number of nodes in $G$.

Summarizing, we have explained the relation between elimination orders and tree decompositions and provided algorithms to map between them in the Appendices \ref{sec:building_tree_decomposition} and \ref{sec:finding_orderings}. We also provided algorithms to calculate treewidth using either tree decomposition or any of its associated elimination orders. The computational complexity of tensor contractions depends on the treewidth corresponding to the given variable elimination order. In the next section, we review the approach of Chen et al.\onlinecite{Chen2018} to parallel tensor contraction.

\subsection{Graphical models in parallel tensor contraction\label{ssec:parallel_tensor_contraction}}
In this section, we focus on an algorithm for parallel tensor network contraction, which is based on graphical models. The "one index at a time" tensor network contraction presented in Sec.~\ref{ssec:graphical_models} (sometimes called Bucket elimination~\cite{detcher2013bucket}) is an inherently sequential operation. Given an elimination order $\pi$, the indices in a tensor network (or a quantum circuit) are removed one-by-one according to $\pi$, and, in general, the elimination of the index with higher-order in $\pi$ can not be performed before all lower-order indices are eliminated. 
In the following, we employ the idea of Chen et al.~\onlinecite{Chen2018} to parallelize the contraction algorithm.

Take as an example the network in Eq.~\ref{eq:tensor_network}. We may choose some index, say $k$, and fix its value within its range. The resulting subnetworks will have one less index than the original expression, as shown in Eq.~\ref{eq:fixing_indices}. Let us denote the result of the contraction of subexpressions, corresponding to different values of the index $k$, as $\sigma_{k}$.   
\begin{equation}
    \begin{split}
        & \sum_{ijklmn} A_{i} B_{ijk} C_{jl} D_{kl} E_{km} F_{ln} G_{mn} = \sigma \\
        & \sum_{ijlmn} A_{i} B_{ij1} C_{jl} D_{1l} E_{km} F_{ln} G_{mn} = \sigma_{1} \\ 
        & \sum_{ijlmn} A_{i} B_{ij2} C_{j2} D_{2l} E_{km} F_{ln} G_{mn} = \sigma_{2} \\
        & \ldots \\
        & \sum_{ijlmn} A_{i} B_{ijL} C_{jl} D_{1l} E_{1m} F_{ln} G_{mn} = \sigma_{L}
    \end{split}
    \label{eq:fixing_indices}
\end{equation}
 It is evident that the result of the contraction of the full expression is equivalent to the sum of contributions from all subexpressions (Eq.~\ref{eq:total_results}).
\begin{equation}
    \begin{split}
         & \sigma = \sum_{k=1}^{L} \sigma_{k}
    \end{split}
    \label{eq:total_results}
\end{equation}
The central point of the described idea is that subexpressions are independent of each other and can be evaluated in parallel. Repeating
the procedure for $m$ variables results in $L^{m}$ independent subtasks.

\begin{figure}[ht!]
\centering
\includegraphics[width=0.48\textwidth]{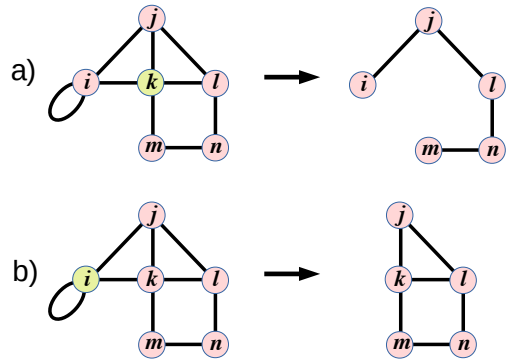}
\caption{Fixing a value of a variable corresponds to vertex removal in a graphical model. The resulting graph represents a subexpression (a subtask), which can be evaluated independently. Removal of different vertices results in subtasks having different complexities. The treewidth of the reduced graph is 1 in case \emph{a)} and 2 in case \emph{b)}.
\label{fig:index_removal}}
\end{figure}

The removal of an index from the initial expression is equivalent to removing the corresponding vertex from the expression's graph, as shown in Fig.~\ref{fig:index_removal}. The resulting reduced graph corresponds to the subexpression with a fixed index. Notice that different choices of the indices for parallelization results in subexpressions (subtasks) of different complexity. The treewidth of the reduced graph characterizes the complexity of the subexpression. 

Consider two choices of indices in Fig.~\ref{fig:index_removal}. In case \emph{a)}, the treewidth of the reduced graph is 1, as the reduced graph is a path graph, while in case \emph{b)}, the treewidth equals 2, as the reduced graph contains a clique on three vertices. In order to find an efficient parallelization scheme, it is imperative to select the vertices for removal such that the treewidth of the resulting subgraph is minimized. This problem is known in graph-theoretic literature as the $\mu$-treewidth deletion problem and is NP-complete\cite{fomin2012planar}.
In the next section, we present several ideas to build efficient heuristics to solve it.

\section{Heuristics for efficient parallel tensor contraction\label{sec:heuristics}}
In previous section we show that in order to implement efficient parallel contraction of tensor networks (or simulation of quantum circuits), one needs to carefully select for removal the vertices of the expression's graph. Suppose we need to remove up to $m$ vertices from the initial graph $G$. We will denote the set of removed vertices by $\mu$. The choice of $\mu$ can be made one vertex at a time based on the maximization of some \textit{score} or \emph{objective} function defined on the vertices of $G$. Careful selection of such function $f: G \rightarrow \mathbb{R}$ is a critical task. One of the criteria for $f$ is low computational cost.

Recall that in the context of tensor contraction or quantum circuit simulation we have access to the tree decomposition of the expression's graph in the form of the elimination order, because we need to find an optimal elimination order anyway to perform the contraction/circuit simulation. This information can be reused while searching for $\mu$. A general greedy algorithm for the $\mu$-treewidth deletion problem is listed in Alg.~\ref{code:greedy_algorithm}. This program takes a graph $G$ and its elimination order and outputs a reduced graph $\tilde{G}$, the set of removed vertices $\mu$ and the elimination order of the reduced graph. In the following we consider different score functions for the greedy algorithm.
\begin{algorithm}[H]
  \caption{Greedy treewidth deletion algorithm}
  \label{code:greedy_algorithm}
  \begin{algorithmic}[1]
    \Require $G = (U, E)$, $\pi = \{(u_{i}, i)\}_{i=1}^{|U|}$, $m$
    \Ensure $\tilde{G}$, $\tilde{\pi} = \{(u_{i}, j)\}_{j=1}^{|U| - m}$, $\tau$
    \Statex
    \Function{Greedy\_treewidth\_deletion}{$G, \pi, m$}
    \State $\mu \gets \emptyset$
    \State $\tilde{G} = G$
    \State $\tilde{\pi} = \pi$
    \For{$j \in [1\dots m]$} \Comment{remove $m$ vertices}
        \State $u^{\ast} = \underset{u \in \tilde{G}}{argmax}(f(\tilde{G}, \tilde{\pi}))$ 
        \State $\tilde{G} \gets \tilde{G} \setminus u^{\ast}$
        \State $\tilde{\pi} \gets \tilde{\pi} \setminus u^{\ast}$
        \State $\mu \gets \mu \cup u^{\ast}$
        \State Optional: $\tilde{\pi} \gets $ tree\_decomposition$(G)$ \label{code:line:td_recalculate}
    \EndFor
    \State $\tau \gets$ find\_treewidth\_from\_order$(\tilde{G}, \tilde{\pi})$
    \EndFunction
  \end{algorithmic}
\end{algorithm}

\subsection{The choice of the score function for greedy algorithm\label{ssec:metrics}}
Different vertex-valued functions can be chosen for a greedy algorithm.
One of the simplest options is the  \emph{degree} function, e.g. the number of neighbors of a vertex. The intuition is the following: removing vertices with the maximal number of neighbors should break large cliques and decrease the treewidth of the resulting graph. We also consider the function based on the \emph{betweenness centrality}, which is the number of shortest paths between all pairs of vertices that go through the chosen vertex. Removing vertices with high centrality makes the graph less connected. These choices, however, do not take into account the information contained in the elimination order.

Another option is to directly minimize the treewidth of the reduced graph, similar to the approach of Chen et al.~\cite{Chen2018}. Given a graph and its elimination order, we test the removal of each vertex, which results in different reduced graphs and corresponding reduced elimination orders (orders where one vertex is removed, but the relative order of the rest of vertices is not changed). The treewidth of the reduced graph is calculated using the reduced elimination order (for example, with Alg.~\ref{code:find_treewidth_order}) and the difference with the initial treewidth is the result of $f(\tilde{G}, \tilde{\pi})$. The treewidth reduction-based score is listed in Alg.~\ref{code:tw_reduction_metric} 

\begin{algorithm}[H]
  \caption{Direct treewidth minimization score}
  \label{code:tw_reduction_metric}
  \begin{algorithmic}[1]
    \Require $G = (U, E)$, $\pi = \{(u_{i}, i)\}_{i=1}^{|U|}$
    \Ensure $u^{\ast}$
    \Statex
    \Function{Direct\_treewidth\_metric}{$G, \pi$}
    \State $\tau \gets$ {Find\_treewidth\_from\_order}($G, \pi$)
    \State $\Delta \gets 0$
    \For{$u \in U$}
            \State $\tilde{\pi} \gets \pi \setminus u$
            \State $\tilde{G} \gets G \setminus u^{\ast}$
            \State $\tilde{\Delta} \gets \tau - $ {find\_treewidth\_from\_order}($\tilde{G}, \tilde{\pi}$)
            \If{$\tilde{\Delta} > \Delta$}
                \State $u^{\ast} \gets u$
            \EndIf
    \EndFor
    \EndFunction
  \end{algorithmic}
\end{algorithm}

Several points should be mentioned about the properties of TDs /elimination orders with respect to vertex removal.

First, notice that the treewidth can be reduced at most by $\Delta = 1$ by removing a single vertex. This fact is evident from the definition of the treewidth. By removing a single vertex from the graph $G$, the size of the maximal bag in the corresponding tree $F$ is reduced by 1. If multiple maximal bags are containing distinct sets of vertices of $G$, then the treewidth will not be reduced by a single vertex removal. As a consequence, the treewidth is a monotonic non-increasing function of the number of removed vertices.

\begin{figure}
\centering
\includegraphics[width=0.48\textwidth]{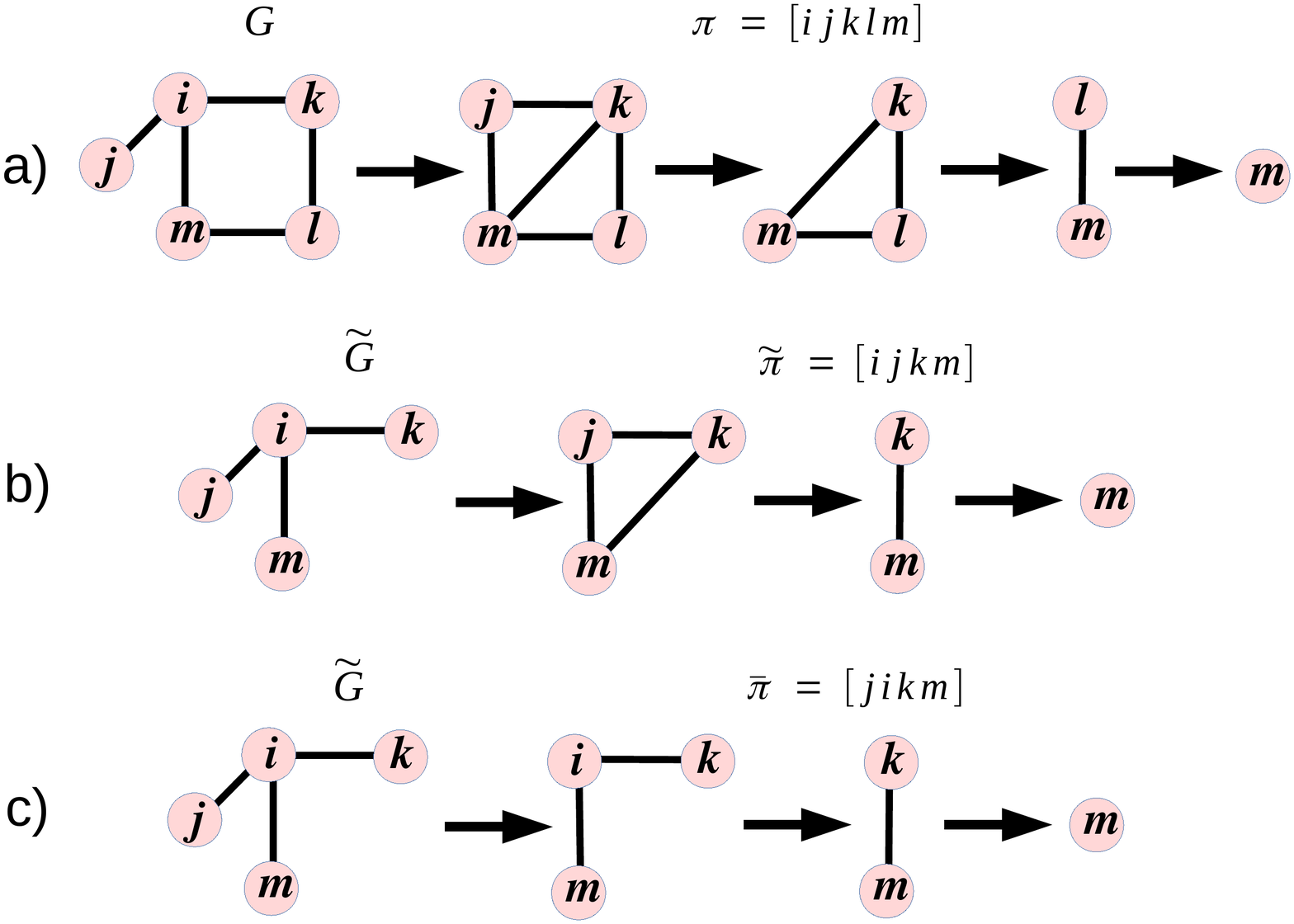}
\caption{Reduced graphs and elimination orders.\\ \emph{a)} Graph $G$ and its optimal elimination order corresponding to treewidth 2.\\ \emph{b)} Reduced graph $\tilde{G}$ and its reduced elimination order $\tilde{\pi}$. The order $\tilde{\pi}$ is not optimal and corresponds to treewidth 2.\\ \emph{c)} Reduced graph $\tilde{G}$ and its optimal elimination order $\bar{\pi}$, which corresponds to treewidth 1.\label{fig:inoptimal_elimination}}
\end{figure}

Another observation is that the elimination order/TD may not remain optimal after removing a vertex from the graph (e.g., the reduced elimination order may not correspond to minimal treewidth). As an example, consider a graph $G$ in Fig.~\ref{fig:inoptimal_elimination} and it's elimination order $\pi$. The treewidth of the reduced graph $\tilde{G}$ is 1, although the reduced-order $\tilde{\pi}$ corresponds to treewidth 2, and the optimal order is $\bar{\pi}$. The provided example shows that the elimination order has to be recalculated several times to remain optimal (line \ref{code:line:td_recalculate} in the Alg.~\ref{code:greedy_algorithm}). 

Without an optimal elimination order, the treewidth reduction-based score quickly fails to find proper deletion set $\mu$. The recalculation of the elimination order entails solving an NP-complete TD problem (or finding an approximate solution) and may be time-consuming. However, if one could obtain the elimination order with the lowest treewidth after removal of each vertex, then the greedy algorithm with the treewidth reduction score would yield the best possible parallelization scheme. 

\subsection{Tree-trimming score\label{ssec:tree_trimming_metric}}
In this section, we introduce a new heuristic score function for the $\mu$-treewidth deletion problem in the general greedy algorithm.
This function employs the information about the TD and yields close to optimal solutions (in the sense of the treewidth of the reduced graph) even after several vertices are removed. The idea of the score is based on the properties of tree decomposition.

\begin{figure}
\centering
\includegraphics[width=0.48\textwidth]{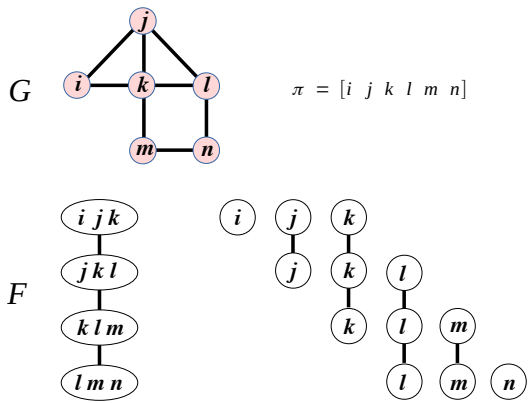}
\caption{The structure of the tree decomposition. The tree $F$ is a TD of the graph $G$ and corresponds to the elimination order $\pi = [i ~ j ~ k ~ l ~ m ~ n]$. Each bag in $F$ has size 3, and the treewidth is 2. The tree $F$ is an intersection of the subtrees of individual vertices of $G$. \label{fig:subtree_decomposition}}
\end{figure}

Recount the third property from the definition of TD: for each vertex $u$ of the graph $G$, the TD $F$ contains a connected subtree. The tree decomposition is thus an intersection of subtrees of the vertices of $G$, which is shown in Fig.~\ref{fig:subtree_decomposition}.
The idea of the proposed heuristic is to pick vertices greedily with respect to the width of $F$ and the shape of the eliminated subtree. Specifically, at each step, the algorithm proceeds as follows: 
\begin{enumerate}
    \item Find the largest bag $b_{max}$ in $F$ (which determines treewidth). If multiple maximal size bags are found, consider their union as $b_{max}$.
    \item For each node $u$ in $b_{max}$, find its weighted subtree $S_{u}$. The weight of each node $v$ in the subtree $S_{u}$ is the size of the bag in $F$ the node $v$ belongs to.
    \item Select the subtree $S^{\ast}_{u}$ with maximal length. In case of equal length subtrees, select the subtree with maximal weight. If the latter condition does not break a tie, then break tie randomly. Return the vertex $u^{\ast}$ corresponding to the selected subtree.
\end{enumerate}

The rationale behind the procedure is natural. The greedy algorithm is guaranteed to reduce the treewidth of the graph provided there is a single largest bag in the TD, and the TD is close to optimal. By removing the longest subtree, we aim to eliminate the "most influential" vertex in the TD. At the same time, we are guaranteed to reduce the treewidth if it is possible since the vertices are removed only from maximal bags. The use of this non-local score $f(G, \pi)$ significantly mitigates the shortcomings of the greedy approach. 
Note that the numerical cost of the score function is polynomial, as it involves only the search in the tree $F$. The proposed score is thus much faster than greedy approaches based on the recalculation of TD.

\section{Numerical experiments\label{sec:experiments}}
In this section, we benchmark parallelization algorithms for the task of quantum circuit simulation. All numerical experiments were performed with our quantum circuit simulation library called "QTree" \cite{schutski2019adaptive}, which is implemented in Python~\cite{oliphant2007python}. We use the NetworkX library to manipulate graphs~\cite{hagberg2008exploring}. To calculate (approximate) TD decompositions, we employed the program of Tamaki et al.~\cite{tamaki2019positive} with execution time constrained to 120 seconds. 

For experiments, we used circuits by Boixo et al.~\cite{boixocircuits}, which are available online. We selected a $7 \times 7$-qubit circuit of depth 50, which results in a tensor network with 723 variables and 1544 tensors.

Provided the initial expression's graph $G$, the parallelization algorithm produces a list of removed nodes (vertices) $\mu$, the reduced graph $\tilde{G}$ (subexpression evaluated in parallel) and the contraction order of the reduced graph $\tilde{\pi}$. 
We tested the dependence of treewidth of the reduced graphs produced by Alg.~\ref{code:greedy_algorithm} for different choices of score function. 
 
\begin{figure}[ht]
\centering
\includegraphics[width=0.5\textwidth]{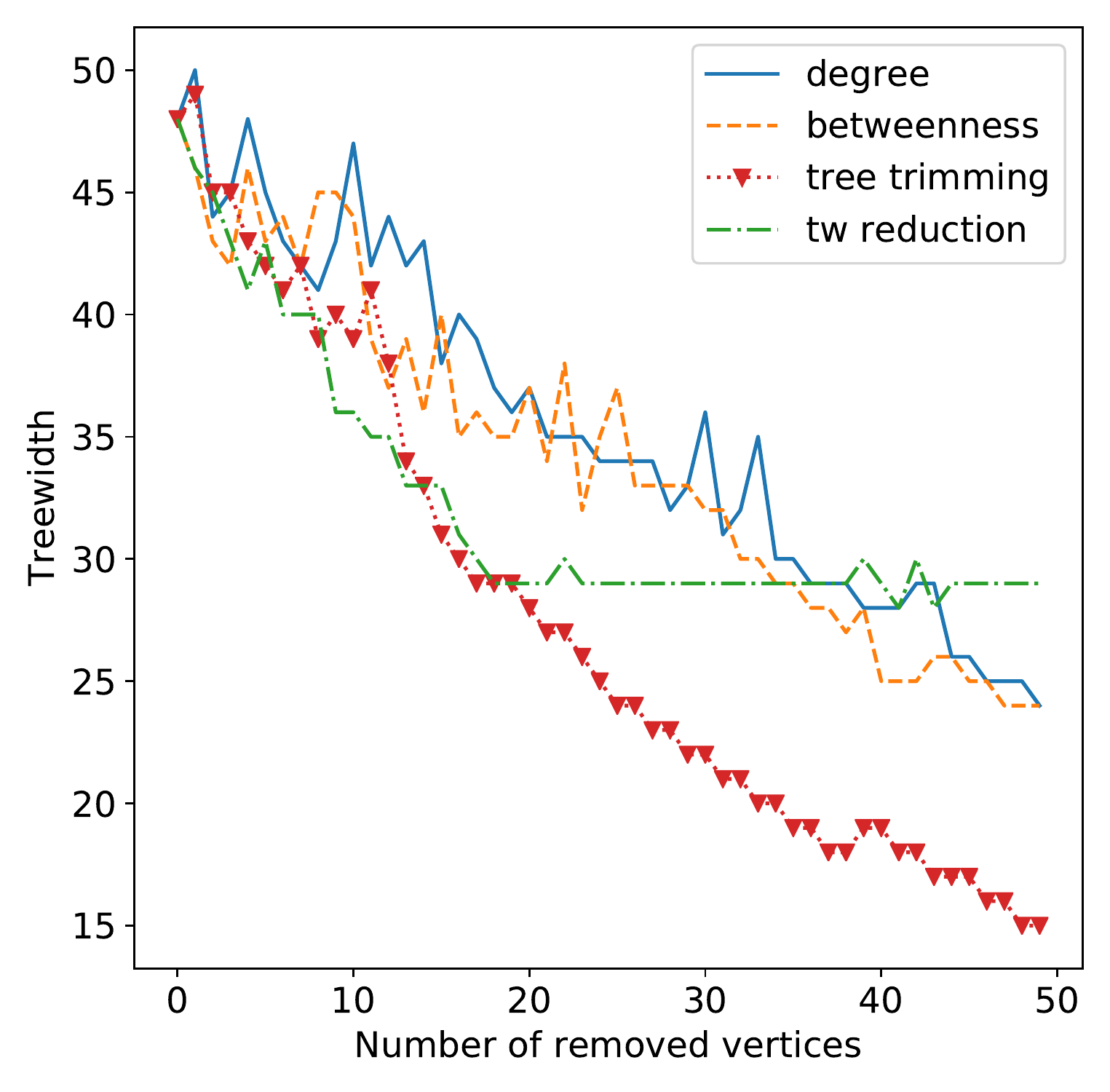}
\caption{Treewidth of the reduced graph for different choices of score function in Alg.~\ref{code:greedy_algorithm}.\label{fig:tw_metric_choice}}
\end{figure}

In the first set of experiments, we run Alg.~\ref{code:greedy_algorithm} without recalculating the TD. The results for different scores are shown in Fig.~\ref{fig:tw_metric_choice} 

We have to note that the non-monotonic behavior of curves of Fig.~\ref{fig:tw_metric_choice} is an artifact of the approximate TD solver. If the solver would be provided large enough execution time (exponential in treewidth), then Fig.~\ref{fig:tw_metric_choice} would show a non-increasing dependence of treewidth on the number of removed vertices (see discussion in the previous subsection). All following figures are indicating either upper or lower bounds of the appropriate quantities, which are found by solving the TD problem with chosen algorithm and computation time budget.  

As shown in Fig.~\ref{fig:tw_metric_choice}, the treewidth-reduction score, as well as our novel tree-trimming score, provide for the fastest decrease of treewidth. However, the treewidth-reduction score has problems after a large number of vertices are removed, and the reduced elimination order $\tilde{\pi}$ is not any more close to an optimal one. In the latter case the greedy algorithm is unable to remove vertices which will lead to the decrease of treewidth. 

\begin{figure}
\centering
\includegraphics[width=0.5\textwidth]{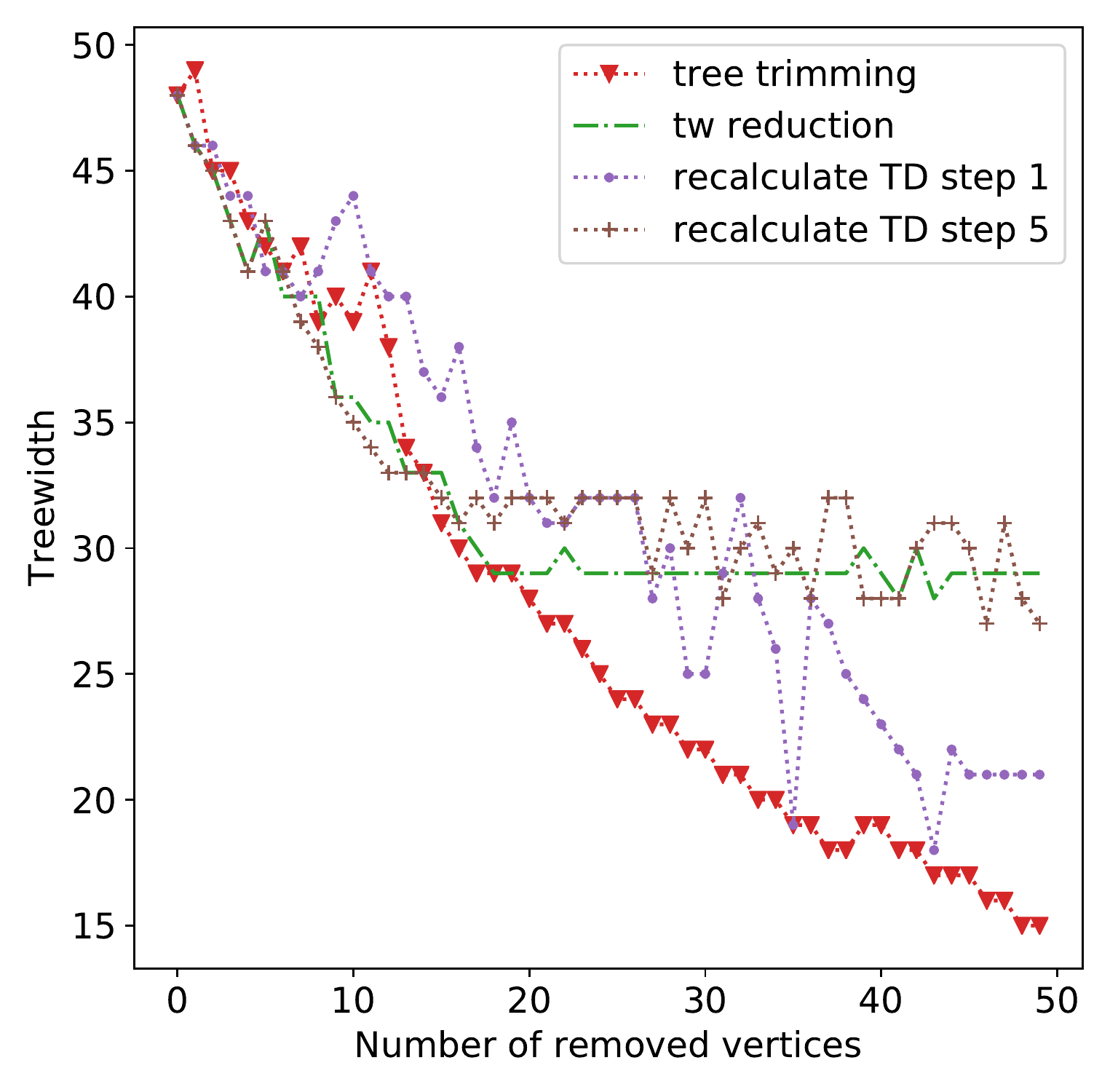}
\caption{Dependence of the performance of the treewidth-reduction score on the frequency of updates of TD. Tree-trimming score is shown for comparison.\label{fig:tw_td_recalculation}}
\end{figure}

In the next set of experiments, we test the effect of TD recalculation on the performance of the score based on the treewidth-reduction. The results are shown in Fig.~\ref{fig:tw_td_recalculation}. If the elimination order is updated frequently (each step), then the approaches based on the treewidth-reduction score sometimes outperforms our heuristic tree-trimming score function. However, if the frequency of updates is not high enough, the treewidth-reduction score quickly results in non-optimal solutions as the size of the deletion set $\mu$ increases.

\begin{figure}
\centering
\includegraphics[width=0.5\textwidth]{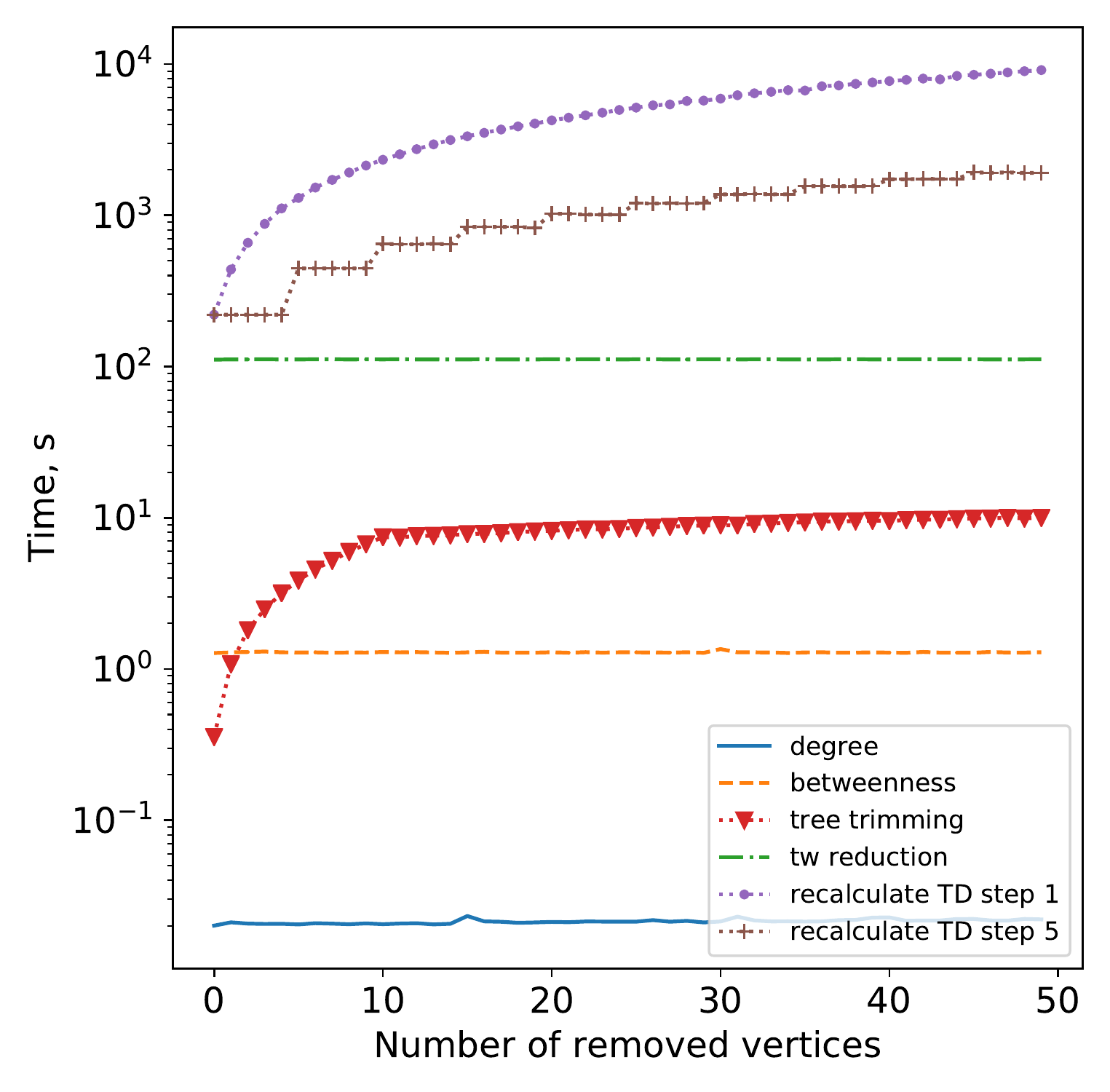}
\caption{Execution time of different $\mu$-treewidth deletion heuristics as a function of the number of removed vertices. Please note the logarithmic scale\label{fig:exec_time}}
\end{figure}

We also provide timings for both experiments in Fig.~\ref{fig:exec_time}. The degree, betweenness, and treewidth-reduction based algorithms evaluate the score only once, and hence the execution time does not depend on the number of removed vertices. The degree and betweenness-based algorithms are the fastest; they use efficient NetworkX built-in implementations of the score function. The execution time of the tree-trimming algorithm grows linearly with the number of removed vertices because the tree-trimming score has to be recalculated after each update to the initial TD. If in Alg.~\ref{code:greedy_algorithm}, the treewidth-reduction score is supplemented by the recalculation of TD after each $k$ vertices are removed, then the execution time grows linearly with slope $R / k$, where $R$ is the time required to recalculate TD. The total time required for the treewidth-reduction score with TD updates becomes significant if a large number of vertices needs to be removed.

\begin{figure}[H]
\centering
\includegraphics[width=0.5\textwidth]{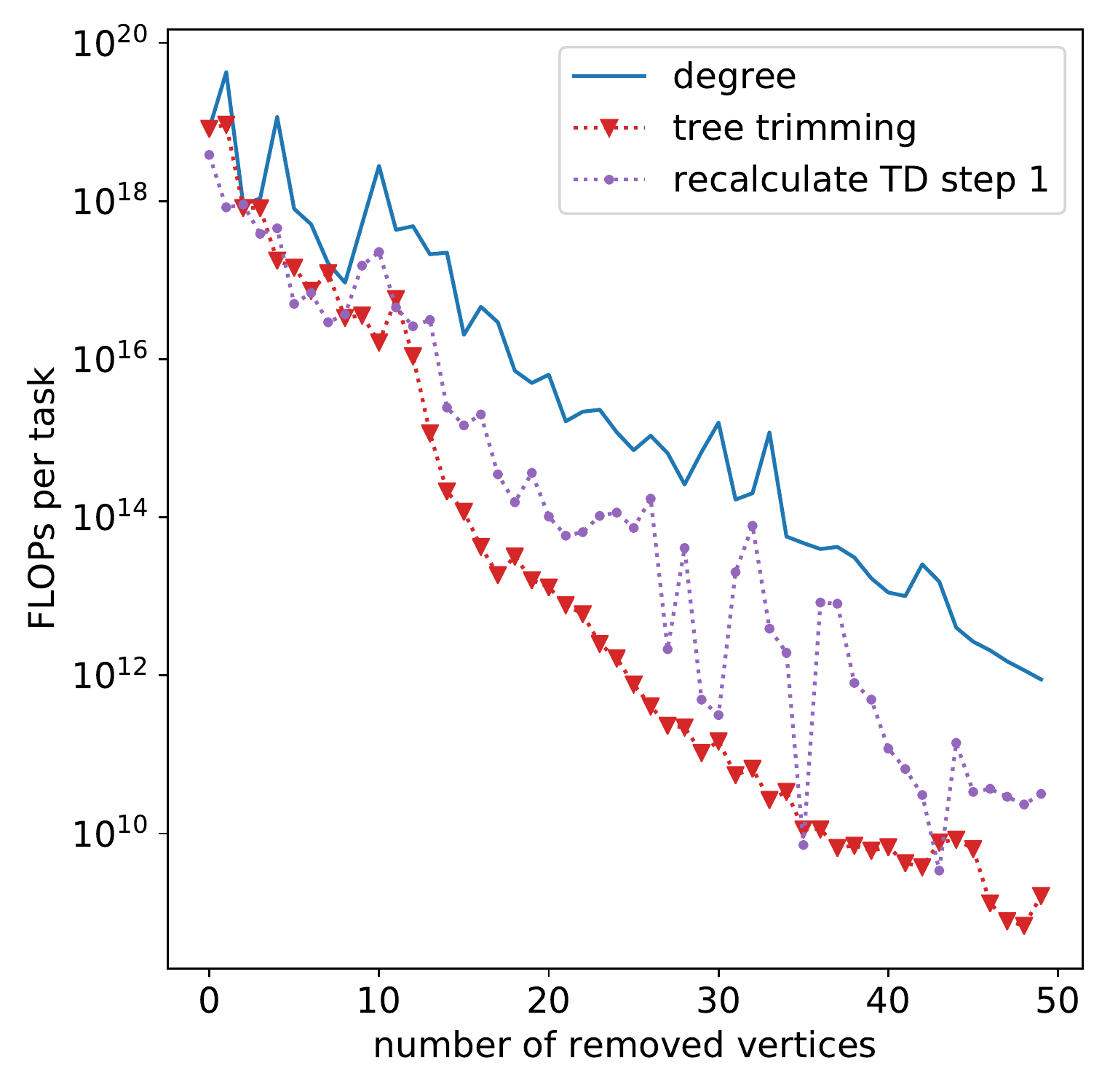}
\caption{Dependence of the numerical effort per task for three selected parallelization heuristics.\label{fig:flops}}
\end{figure}

\begin{figure}[H]
\centering
\includegraphics[width=0.5\textwidth]{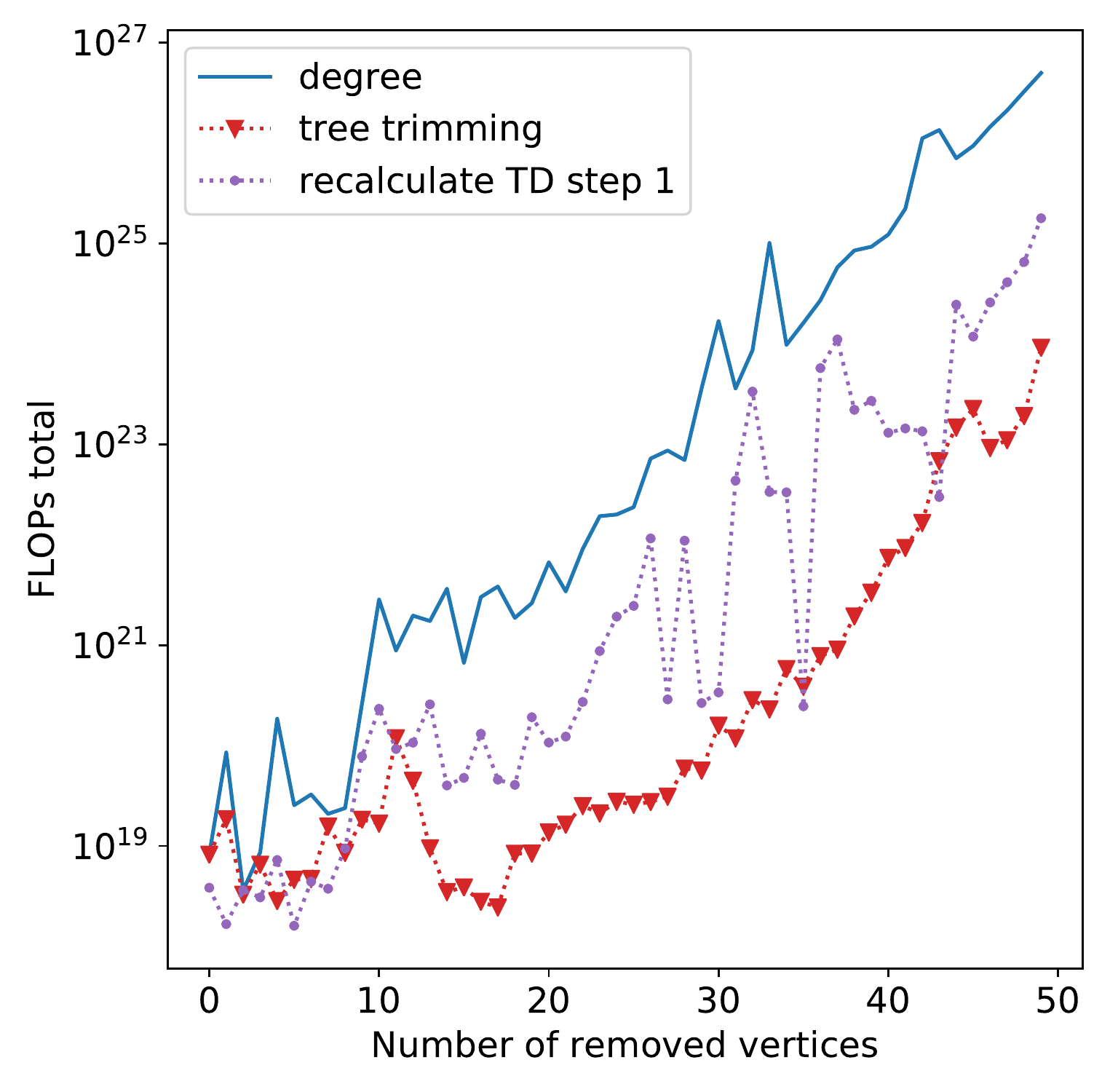}
\caption{Dependence of the total numerical effort for three selected parallelization heuristics.\label{fig:flops_total}}
\end{figure}

\begin{figure}[H]
\centering
\includegraphics[width=0.5\textwidth]{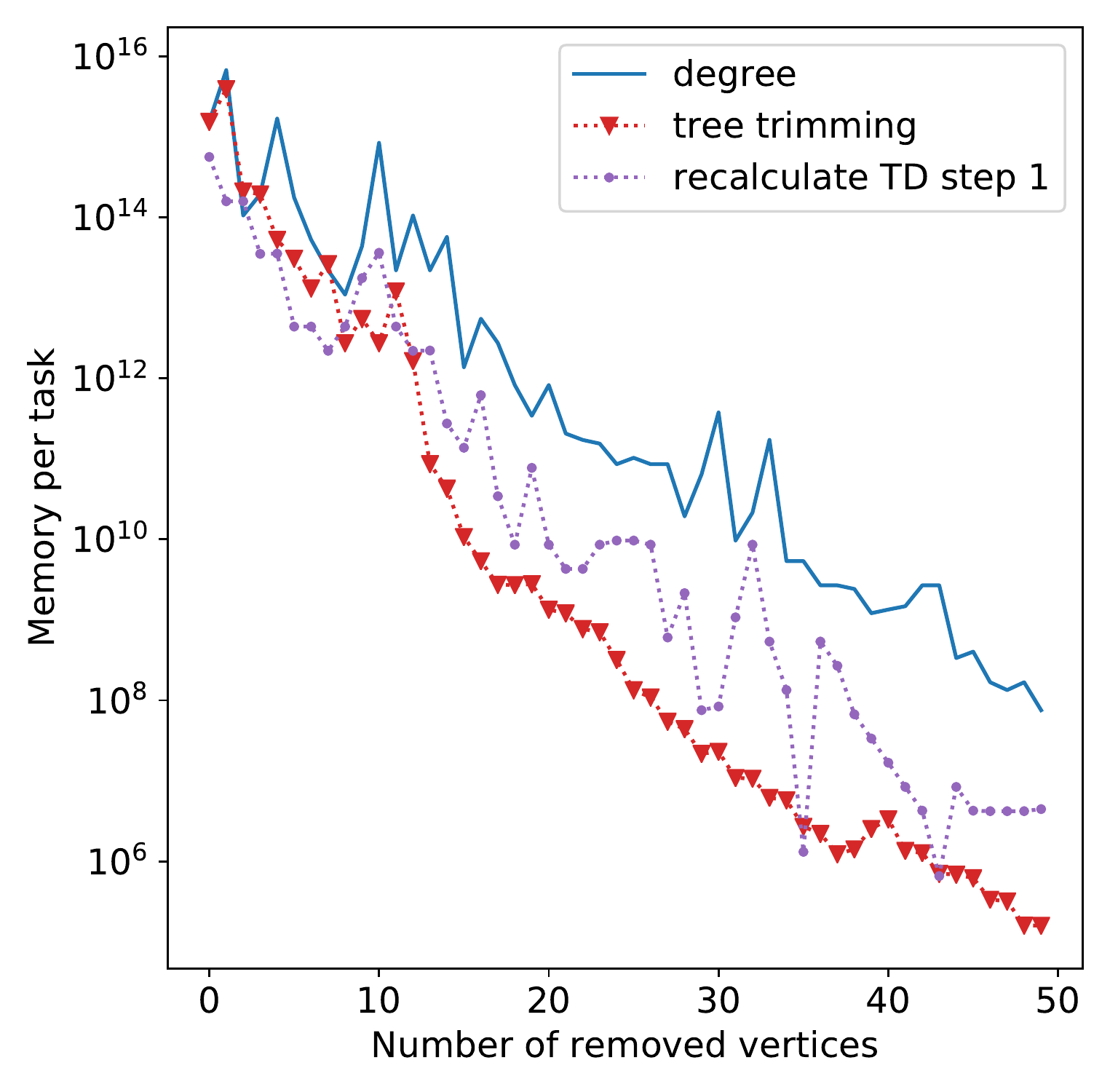}
\caption{Memory per subtask as predicted by different $\mu$-treewidth deletion heuristics.\label{fig:mem}}
\end{figure}

Finally, given the expression's graph and the contraction order, it is possible to calculate the memory requirements and the number of floating-point operations (FLOPs) needed to perform the contraction of the tensor network, for details see Ref. \onlinecite{schutski2019adaptive}. We provide the dependence of FLOPs for each subtask and total FLOPS in Fig.~\ref{fig:flops} and Fig.~\ref{fig:flops_total}. The minimal memory requirement for a subtask is plotted in Fig.~\ref{fig:mem}. We provide results for the worst (degree) and two best heuristics we found (treewidth-reduction with recalculation of TD and tree-trimming).

 The memory $\mathcal{M}$ and FLOPs $\mathcal{P}$ requirements of subtasks depend exponentially on the treewidth $\tau$ of the reduced graph $\tilde{G}$. Specifically, for quantum circuits the dependence is $\mathcal{M} = O(2^{\tau})$ and $\mathcal{P} = O(2^{\tau+1})$ \cite{schutski2019adaptive}. At the same time, total FLOPs required for all subtasks combined is $2^{|\mu|} \times \mathcal{P}$, where $\mu$ is the set of removed vertices.
 
 Exponential scaling of resources with treewidth and the number of parallelized variables highlights the importance of efficient treewidth-reducing heuristics. The simulation of large quantum circuits during the race for "quantum supremacy" is essentially a memory-bound task.\cite{chen201864} Maximal difference in the memory size needed for each subtask between best and worst algorithms in our experiments is around $10^4$. 

\section{Conclusions and outlook\label{sec:outlook}}
In this paper, we formulated the task of parallel tensor network contraction/quantum circuit simulation in the framework of graphical models. Efficient parallelization scheme of tensor network contraction amounts at solving a $\mu$-treewidth deletion problem. We examined different variants of the greedy algorithm and proposed a novel tree-trimming score, which has an advantage in accuracy and execution time compared to other score function choices. We hope our approach will promote the study of algorithm complexity with the help of graphs.
Multiple extensions of the current method are possible. A rigorous application of the algorithm for general tensor network contraction is proposed for future work. Also, the accounting for the parameters of the computational system in the algorithm, such as communication cost or memory locality, is highly desirable. Finally, we leave the possibility for the existence of a more efficient way to extract the solution of the $\mu$-treewidth deletion problem from TD. We hope that our work will promote the $\mu$-treewidth deletion problem in the graph-theoretic community. 

\subsection*{Acknowledgements}
This research was partially supported by the Ministry 
of Education and Science of the Russian Federation (grant 14.756.31.0001). D.K. is supported by Huawei. We performed calculations on the "Zhores"\cite{zacharov2019zhores} supercomputer at Skoltech.   

\bibliographystyle{named}
\bibliography{references}

\appendix
\section{Building tree decomposition\label{sec:building_tree_decomposition}}
Let us now provide a procedure to build a tree decomposition given a specific elimination sequence, or ordering $\pi$. This procedure performs a sequence of contractions and builds a decomposition along the way. The algorithm is our compilation of known results, and analogous algorithms can be found, for example, in \onlinecite{blair1993introduction}. 

The outcome of the algorithm is a rooted tree, so some additional definitions are needed. A rooted tree is a tree where a single vertex $r$ is selected to be root, which allows us to define the parent/child relation. For any node $b$ in the tree, its parent is a first node $p$ on the unique path from $b$ to $r$. In the following we denote a function \textbf{parent}($b$) which returns a parent of a vertex $b$. Likewise, for any node $b$ in the tree its children are all vertices adjacent to $b$ except its parent, e.g. $c: c \in \mathcal{N}(b), c \neq$ \textbf{parent}($b$). Finally, leaves are vertices that have no children. The algorithm is listed in Alg~\ref{code:build_td}.

\begin{algorithm}[H]
  \caption{Building tree decomposition from the elimination order}
  \label{code:build_td}
  \begin{algorithmic}[1]
    \Require $G = (U, E), \pi: U \rightarrow N, ~~ \pi = \{(u_{i}, i)\}_{i=1}^{|U|}$
    \Ensure $F = (B, T)$
    \Statex
    \Function{Build\_clique\_tree}{$G, \pi$}
    \State orphan\_bags $\gets \emptyset$
    \For{$i \in [1, \ldots, |U| - 1]$}
        \State $u \gets \pi^{-1}(i)$
        \For{$w, x \in \mathcal{N}(u)$}\Comment{form a clique}
            \State $E \gets E \cup (w, x)$
        \EndFor
        \If{$\mathcal{N}(u) \neq \emptyset$}
            \State $b = \mathcal{N}(u) \cup u$
        \EndIf
        \State $U \gets U \setminus u$ \Comment{eliminate the node}
        \State drop\_bag $\gets$ \textbf{False}
        \For{$l ~ \text{in orphan\_bags} $}
        \Comment{keep only maximal cliques}
            \If{$b\subset l$}
              \State $b \gets l$
              \State drop\_bag $\gets$ \textbf{True}
              \State break
            \EndIf
        \EndFor
        \For{$l ~ \text{in orphan\_bags} $} \Comment{update the list of orphans}
            \If{$u \in l \cap b$ \textbf{and} $b \not\subset l$} \Comment{add parent}
                \State orphan\_bags $\gets$ orphan\_bags $\setminus ~ l$
                \State $B \gets B \cup b$
                \State $T \gets T \cup (l, b)$
            \EndIf
        \EndFor
        \If{\textbf{not} drop\_bag} \Comment{add leaf to the tree}
            \State $B \gets B \cup b$
        \EndIf
    \EndFor
    \EndFunction
  \end{algorithmic}
\end{algorithm}

For a given order $\pi$, the algorithm performs a sequence of edge contractions. At each step, a clique that contains the next vertex in $\pi$ is added as a new node to the tree. Here we omit cliques, which are subsets of larger cliques: nothing is added to the tree in this case. Thus only \emph{maximal} cliques are kept.

The algorithm builds the tree from the bottom to the root in a breadth-first search way. First, leaf cliques are found. In the next steps, successive layers of parent cliques are added until the root is reached. The list of orphan cliques is stored to find the next layer of parents. A candidate clique is checked against this list. The candidate is a parent of an orphan if the current node in $\pi$ lies in the intersection of the candidate clique with the child clique. We delete the candidate from the list of orphans in this case. Otherwise, the candidate is added to the list of orphans.

\section{Finding elimination orders from TD\label{sec:finding_orderings}}
To complete the discussion we provide an algorithm for finding \emph{some} elimination ordering which is consistent with a given tree decomposition.

The algorithm works on rooted trees. First, an arbitrary bag in the tree $F$ should be selected as root. The elimination order $\pi$ is built starting from the leaves. At each step, a leaf is found, and its parent (if any) is identified. The vertices in the difference between the current leaf bag $b$ and its parent bag $p$ can be added in any order to $\pi$. After all nodes in the difference are added to $\pi$, the leaf is removed. The algorithm is listed in Alg.~\ref{code:build_order}.

\begin{algorithm}[H]
  \caption{Recovering order from the tree decomposition}
  \label{code:build_order}
  \begin{algorithmic}[1]
    \Require $F = (B, T)$
    \Ensure $\pi: U \rightarrow N, ~~ \pi = \{(u_{i}, i)\}_{i=1}^{|U|}$
    \Statex
    \Function{Recover\_elimination\_order}{$F$}
    \State $i = 1$ \Comment{order counter}
    \State root $\gets$ select any $b \in B$ 
    \Statex \Comment{vertices in root will be last in $\pi$} 
    \While {$b \neq$ root}
    \For{$b \in B$}\Comment{Find next leaf}
        \If{$|\mathcal{N}(b)| \leq 1 ~ \text{\textbf{and}} ~ b \neq root$}
            \State \textbf{break}            
        \EndIf
    \EndFor
    \State $p \gets \mathcal{N}(b) \setminus b$ \Comment{Find parent of the leaf}
    \State $m \gets b \setminus p$ \Comment{add nodes from the difference with parent to the order}
    \For{$u \in m$}
        \State $\pi \gets \pi \cup (u, i)$
        \State $i \gets i - 1$
    \EndFor
    \If{$b \neq root$}
        \State $B \gets B \setminus b$
    \EndIf
    \EndWhile
    \For{$u \in root$} \Comment{Add nodes from root}
        \State $\pi \gets \pi \cup (u, i)$
        \State $i \gets i - 1$
    \EndFor
    \EndFunction
  \end{algorithmic}
\end{algorithm}

\end{document}